\documentclass[aps,prl,twocolumn,amsmath,amssymb,footinbib,showpacs,longbibliography,superscriptaddress]{revtex4-1}

\newcommand{\Jnature}{Nature (London)}
\newcommand{\Jnatphys}{Nat. Phys.}
\newcommand{\Jnatphot}{Nat. Photonics}
\newcommand{\Jnatcomm}{Nat. Comm.}

\newcommand{\Jscience}{Science}

\newcommand{\Jpnas}{Proc. Nat.l Acad.  Sci.}

\newcommand{\Jprl}{Phys. Rev. Lett.}

\newcommand{\Jpra}{Phys. Rev. A}
\newcommand{\Jprb}{Phys. Rev. B}

\newcommand{\Jpre}{Phys. Rev. E}
\newcommand{\Jrmp}{Rev. Mod. Phys.}

\newcommand{\Jepl}{Europhys. Lett.}
\newcommand{\Jnjp}{New J. Phys.}

\newcommand{\Jjetp}{Sov. Phys. JETP}

\newcommand{\JRepProgPhys}{Rep. Prog. Phys.}

\newcommand{\JjphysB}{J. Phys. B: At. Mol. Opt. Phys.}
\newcommand{\JjphysC}{J. Phys. C: Solid State Phys.}

\newcommand{\Jadvphys}{Adv. Phys.}

\usepackage{times}
\usepackage[english]{babel}
\usepackage{latexsym}
\usepackage{graphics}
\usepackage{subfigure}
\usepackage{epsfig}
\usepackage{color}
\usepackage{hyperref}
\usepackage{braket} 
\usepackage[T1]{fontenc}
\usepackage[latin9]{inputenc}
\usepackage{amstext}
\usepackage{amssymb}
\usepackage{graphicx}
\usepackage{esint}
\usepackage{braket}
\usepackage{babel}
\usepackage{amsmath}
\usepackage{siunitx}

\hypersetup{
colorlinks=true,
citecolor=blue,
linkcolor=red,
urlcolor=black
}


\newcommand{\e}{\textrm{e}}
\newcommand{\ie}{i.e.}

\newcommand{\etal}{\textit{et al.}}

\newcommand{\IPR}{\textrm{IPR}}
\newcommand{\Er}{E_{\textrm{r}}}
\newcommand{\Vc}{V_{\textrm{c}}}
\newcommand{\kB}{k_{\textrm{\scriptsize {B}}}}
\newcommand{\fs}{f_{\textrm{s}}}

\newcommand{\vect}[1]{{\mathbf{#1}}}
\newcommand{\rr}{\mathbf{r}}
\newcommand{\dd}{\mathrm{d}}
\DeclareMathOperator{\Tr}{Tr}

\newcommand{\asc}{a_{\textrm{\tiny 3D}}}
\newcommand{\aho}{l_\perp}
\newcommand{\atwod}{a_{\textrm{\tiny 2D}}}
\newcommand{\aTwoD}{a_{\textrm{\tiny 2D}}}
\newcommand{\athreed}{a_{\textrm{\tiny 3D}}}
\newcommand{\tildeg}{\tilde{g}}
\newcommand{\tildegnot}{\tildeg_0}

\newcommand{\muc}{\mu_\textrm{c}}
\newcommand{\nc}{n_\textrm{c}}

\newcommand{\lettersection}[1]{\paragraph*{#1.---}}

\begin{document}

\title{
Strongly-Interacting Bosons in a Two-Dimensional Quasicrystal Lattice
}

\author{Ronan Gautier}
\affiliation{CPHT, CNRS, Ecole Polytechnique, IP Paris, F-91128 Palaiseau, France}

\author{Hepeng Yao}
\affiliation{CPHT, CNRS, Ecole Polytechnique, IP Paris, F-91128 Palaiseau, France}

\author{Laurent Sanchez-Palencia}
\affiliation{CPHT, CNRS, Ecole Polytechnique, IP Paris, F-91128 Palaiseau, France}

\date{\today}

\begin{abstract}
Quasicrystals exhibit exotic properties inherited from the self-similarity of their long-range ordered, yet aperiodic, structure.
The recent realization of optical quasicrystal lattices paves the way to the study of correlated Bose fluids in such structures, but the regime of strong interactions remains largely unexplored, both theoretically and experimentally. Here, we determine the quantum phase diagram of two-dimensional correlated bosons in an eightfold quasicrystal potential. Using large-scale quantum Monte Carlo calculations, we demonstrate a superfluid-to-Bose glass transition and determine the critical line. Moreover, we show that strong interactions stabilize Mott insulator phases, some of which have spontaneously broken eightfold symmetry.
Our results are directly relevant to current generation experiments and, in particular, drive prospects to the observation of the still elusive Bose glass phase in two dimensions and exotic Mott phases.
\end{abstract}

\maketitle

Quasicrystals are a fascinating state of matter, characterized by long-range, although nonperiodic, order. Such exotic structures may be realized by the continuous tiling of space using sets of irreducible unit cells arranged aperiodically~\cite{penrose1974,gruunbaum1987} or as incommensurable projections of periodic lattices in higher dimensions~\cite{senechal1995}.
Such structures spontaneously appear in the growth of certain alloys~\cite{shechtman1984,steuer2004,steurer2018} or can be engineered in photonic~\cite{valy2013,tanese2014,barboux2017,goblot2020} and ultracold-atom~\cite{lsp2010,modugno2010,mace2016} systems.
The hallmark of quasicrystalline order, \ie\ sharp spots in reciprocal space with a rotation symmetry incompatible with discrete translation invariance, can then be characterized using Bragg~\cite{shechtman1984,levine1984} or matterwave~\cite{lsp2005,viebahn2019} diffraction.
Quasicrystals exhibit unique properties, inherited from
structural self-similarity at all scales. It includes
nontrivial topological order~\cite{lang2012,kraus2012,kraus2013,darreau2017,huang2018},
Anderson-like localization~\cite{aubry1980,roati2008},
as well as fractal properties of wave functions~\cite{sutherland1987,yuan2000}, energy spectrums~\cite{kohmoto1983,tang1986,tanese2014,cubitt2015,yao2019}, and phase diagrams~\cite{yao2020}.
So far, quasicrystals have been extensively studied in regard to
solid-state physics~\cite{shechtman1984,steuer2004,steurer2018}, superconductivity~\cite{kamiya2018}, twisted bilayer graphene~\cite{ahn2018,yao2018b}, photonic structures~\cite{chan1998,lahini2009,freedman2006,vay2013}, and ultracold quantum gases~\cite{guidoni1997,roth2003,damski2003,lsp2005,roati2008,roscilde2008,roux2008,gadway2011,jagannathan2013,lellouch2014,derrico2014,schreiber2015,gori2016,khemani2017,kohlert2019,viebahn2019,sbroscia2020,an2020,yao2020}.

Quasiperiodic Bose fluids are particularly appealing due to the complex interplay of interactions, localization, and quasiperiodicity.
Controlled quasicrystal potentials for atomic systems, free of defects and phonons, can be optically designed using laser fields arranged in various rotation symmetries~\cite{guidoni1997,lsp2005,jagannathan2013}.
Alternatively, quasicrystals can be engineered using long-range interactions, spin-orbit coupling, and cavity-mediated interactions~\cite{gopalakrishnan2013,hou2018,mivehvar2019}.
Moreover interactions can be tuned in wide ranges~\cite{lewenstein2007,bloch2008,pollack2009}, hence offering a unique playground.
Ultracold atoms in one-dimensional (1D) quasiperiodic potentials have been extensively studied in the context of Anderson localization~\cite{roth2003,damski2003,roati2008,an2020},
Bose glasses (BGs)~\cite{roth2003,damski2003,roscilde2008,roux2008,gadway2011,derrico2014,gori2016,yao2020}, and collective~\cite{lellouch2014} and many-body~\cite{schreiber2015,khemani2017,kohlert2019} localization.
In contrast, much less is known in higher dimensions.
Recently, the emergence of quasiperiodic order~\cite{viebahn2019} and localization of
weakly interacting bosons~\cite{sbroscia2020} in a two-dimensional (2D) eightfold quasicrystal have been reported. The existence of a BG and the regime of strong interactions, however, remains largely unexplored.
On the theoretical side, mean field phase diagrams have been found using inhomogeneous Gutzwiller-like ansatz on simplified quasiperiodic graphs~\cite{johnstone2019,johnstone2020,ghadimi2020}.
Such approaches are, however, mean field in nature and ignore the emergence of strong correlations close to critical points  as well as a realistic connectivity of optical quasicrystals.

In this Letter, we study correlated 2D bosons in an eightfold rotationally symmetric quasicrystal potential. Using path integral Monte Carlo calculations, we find exact quantum phase diagrams, taking into account possibly strong interactions and the full quasicrystalline structure of the potential. For weak interactions, we find a superfluid (SF) and a BG phase, determined by the competition of interactions and localization. The SF order parameter shows a clear critical behavior, from which we extract the critical line in the interaction-quasicrystal amplitude diagram. In contrast, the compressibility shows a smooth crossover that is consistent with the compressible character of both phases.
For strong-enough interactions, Mott lobes open within the BG phase due to the competition of particle repulsion, localization, and tunneling.
In most cases, the total filling is a multiple of $8$ which is consistent with the eightfold rotation symmetry of the potential. In some cases, however, we find a multiply-degenerated ground state manifold characterized by spontaneously broken rotation symmetry.
We attribute this behavior to the suppression of double occupancy in pairs of nearby potential wells.
Finally, we discuss experimental and theoretical prospects.

\lettersection{Model and single-particle properties}
We consider a 2D gas of $N$ interacting bosons, of mass $m$, in a quasicrystal potential $V(\rr)$. It is governed by the Hamiltonian
\begin{equation}\label{eq:Hamiltonian}
  \hat{H} = \sum_j \left[ -\frac{\hbar^2}{2m} \nabla^2_j + V(\hat{\rr}_j) \right] + \sum_{j<k} U(\hat{\rr}_j - \hat{\rr}_k)
\end{equation}
where $\hat{\rr}_j$ is the position of the $j$ th particle and $U$ is a short-range repulsive two-body interaction term.
The quasicrystal potential is chosen to be eightfold rotation symmetric and centered on $\rr=0$,
\begin{equation}\label{eq:QPpotential}
V(\rr) = V_0 \sum_{k=1}^4 \cos^2 \left(\vect{G}_k \cdot \rr \right),
\end{equation}
where $V_0$ is the potential amplitude and the quantities $\vect{G}_k$ are the lattice vectors of four mutually incoherent standing waves oriented at the angles \ang{0}, \ang{45}, \ang{90}, and \ang{135}, respectively.
The lattice vectors have norm $|\vect{G}_k| = \pi/a$. We use the lattice spacing $a$ and the corresponding recoil energy $\Er = \pi^2\hbar^2/2ma^2$
as the space and energy units, respectively~\cite{note:SupplMat}.
The eightfold quasiperiodic potential~(\ref{eq:QPpotential}) has been recently realized for a system of ultracold bosons in Refs.~\cite{viebahn2019,sbroscia2020}.

The single-particle properties of the shallow 2D quasicrystal potential are qualitatively similar to its 1D counterpart (see, for instance, Refs.~\cite{boers2007,biddle2009,biddle2010,yao2019}).
The critical localization potential $\Vc$ is found from accurate finite-size scaling analysis of the inverse participation ratio for the single-particle ground state ($\IPR_0$), using exact diagonalization~\cite{note:SupplMat}.
It shows a sharp transition between an extended phase for $V<\Vc$ and a localized phase for $V>\Vc$. Using square quasicrystal lattices of linear sizes up to $L=128a$, we find $\Vc/\Er \simeq 1.76 \pm 0.01$ which is in good agreement with the result of Ref.~\cite{szabo2020} found using another approach (ground-state curvature).
Moreover, we find the critical behavior $\IPR_0 \sim (V-\Vc)^\nu$ with the universal exponent $\nu \simeq 1/3$~\cite{note:SupplMat}.

We now turn to interacting bosons.
In the low-energy $s$-wave scattering limit, the interaction potential $U(\rr)$ is fully characterized by the 2D scattering length $\atwod$.
In practice, a quasi-2D Bose gas may be realized by strongly confining a 3D gas to zero point transverse oscillations. For a harmonic trap of angular frequency $\omega_\perp$, it requires that the excitation energy exceed the chemical potential $\mu$ and the temperature $T$, $\hbar\omega_\perp \gg \mu, \kB T$,
with $\kB$ the Boltzmann constant.
The 2D scattering length is then determined by its 3D counterpart and the characteristic transverse length $\aho = \sqrt{\hbar/m\omega_\perp}$~\cite{petrov2000a,petrov2001},
\begin{equation}\label{eq:a2d-a3d}
\atwod \simeq 2.092\aho \mathrm{exp}\left( -\sqrt{\frac{\pi}{2}} \frac{\aho}{\asc} \right).
\end{equation}
On the other hand, the interaction strength is characterized by the dimensionless mean field coupling constant $\tildeg$, such that the energy per particle in the homogenous gas is $E/N = \tildeg \times (\hbar^2 n/2m$), with $n$ the 2D density.
In 2D, the quantity $\tildeg$ depends not only on $\atwod$ but also on the chemical potential $\mu$~\cite{petrov2000a,petrov2001,pricoupenko2007,bloch2008}.
Up to logarithmic accuracy, it may be conveniently written~\cite{note:SupplMat}~
\begin{equation}\label{eq:coupling}
\tildeg \simeq \frac{1}{\tildegnot^{-1} + (4\pi)^{-1}\ln\left(\Lambda \Er/\mu\right)},
\end{equation}
where $\Lambda \simeq 0.141$ is a numerical constant and
\begin{equation}\label{eq:coupling0}
\tildegnot = \frac{2\pi}{\ln(a/\atwod)}.
\end{equation}
The quantity $\tildegnot$ is the relevant interaction parameter we shall use in the following.

\lettersection{Weak interactions}\label{sec:sec3}
We start with weakly interacting bosons, $\tildeg \ll 1$.
In this regime, the phase diagram results from the competition of localization and interactions, and the superfluid fraction $\fs$ may serve as an order parameter.
While the quasicrystal potential tends to localize the bosons for $V>\Vc$ and favor a BG phase ($\fs=0$), the repulsive interactions tend to delocalize the bosons and restore superfluidity (SF, $\fs>0$).

To determine the phase diagram, we first use a mean field approach.
The ground-state dynamics of the Bose gas is then governed by the Gross-Pitaevskii equation (GPE)
\begin{equation}\label{eq:GPE}
\mu \psi = {-\hbar^2 \vect{\nabla}^2} \psi / {2m} + V(\vect{r}) \psi + g N \vert\psi\vert^2 \psi
 \end{equation}
where $\psi(\vect{r})$ is the classical field and $g=(\hbar^2/m)\tildeg$ the dimensionful coupling constant,
and we use the normalization condition $\int d\vect{r} |\psi(\vect{r})|^2 = 1$.
Within the GPE, the mean field phase diagram is determined by only two universal dimensionless parameters, namely the potential amplitude $V_0/\Er$ and the coupling coefficient $gn/\Er$.
The field $\psi(\vect{r})$ is found using imaginary time evolution from an arbitrary state.
It yields the total energy $E$ and the chemical potential $\mu$.
The superfluid fraction is found from the boost approach using twisted boundary conditions,
\begin{equation}\label{eq:superfluid}
\fs = \frac{2m}{\hbar^2 n} \lim_{\Theta \rightarrow 0}\frac{E_\Theta - E_0}{\Theta^2},
 \end{equation}
where $E_\Theta$ is the energy for the phase difference $\Theta$ at opposite sides of the system~\cite{fisher1973,krauth1991}.

 \begin{figure}[t!]
         \centering
         \includegraphics[width = 1.0\columnwidth]{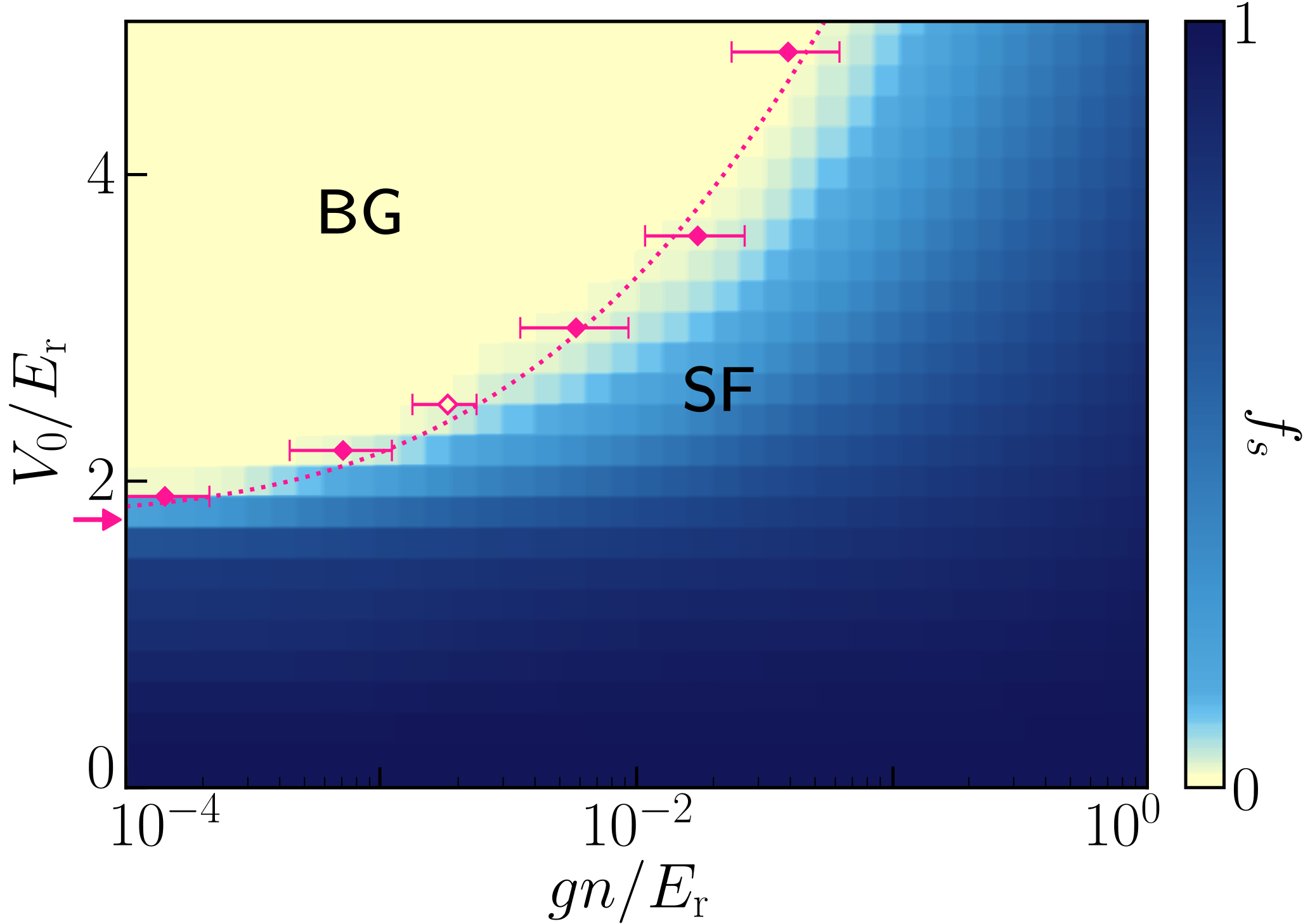}
\vspace{-0.5cm}
\caption{\label{fig:fig1}
Phase diagram of the weakly interacting Bose gas in a 2D quasicrystal lattice.
The mean field SF fraction $\fs$ is shown in color scale for a system of linear size $L=20a$. It exhibits a BG phase ($\fs=0$, yellow) and a SF phase ($\fs>0$, blue), separated by a narrow intermediate region. The exact critical line is found from QMC calculations at $\tildegnot=0.03$
(pink points; the dotted line is a guide to the eye).
The hollow pink point corresponds to the transition found in Fig.~\ref{fig:fig3}.
The pink arrow indicates the single-particle critical point, $\Vc \simeq 1.76\Er$.
}
 \end{figure}

Figure~\ref{fig:fig1} is the phase diagram of the weakly interacting Bose gas
against $V_0/\Er$ and $gn/\Er$. It shows two distinct regimes, separated by a sharp line (see details below). For low interactions and/or a strong quasicrystal potential, the mean field SF fraction vanishes and we find a BG phase (yellow region). Up to numerical accuracy, we find $\fs=0$, except close to the separation line. For strong enough interactions and low quasiperiodic potential, we find $\fs > 0$, corresponding to the SF phase (blue region). For $V_0<\Vc$, there is no localization and the Bose gas is always in the SF phase, although with $\fs<1$, except in the limit of a vanishing quasicrystal potential, $V_0 \rightarrow 0$. As expected, the SF-BG transition coincides with the single-particle localization point in the limit of vanishing interactions, $gn \rightarrow 0$ (pink arrow).
Increasing repulsive interactions compete with localization induced by the quasicrystal potential and the critical point is shifted toward higher potential strengths. An analogous effect has also been observed in 1D interacting Fermi gases~\cite{pilati2017}.

\begin{figure}[t!]
         \centering
         \includegraphics[width = \columnwidth, height=0.8\columnwidth]{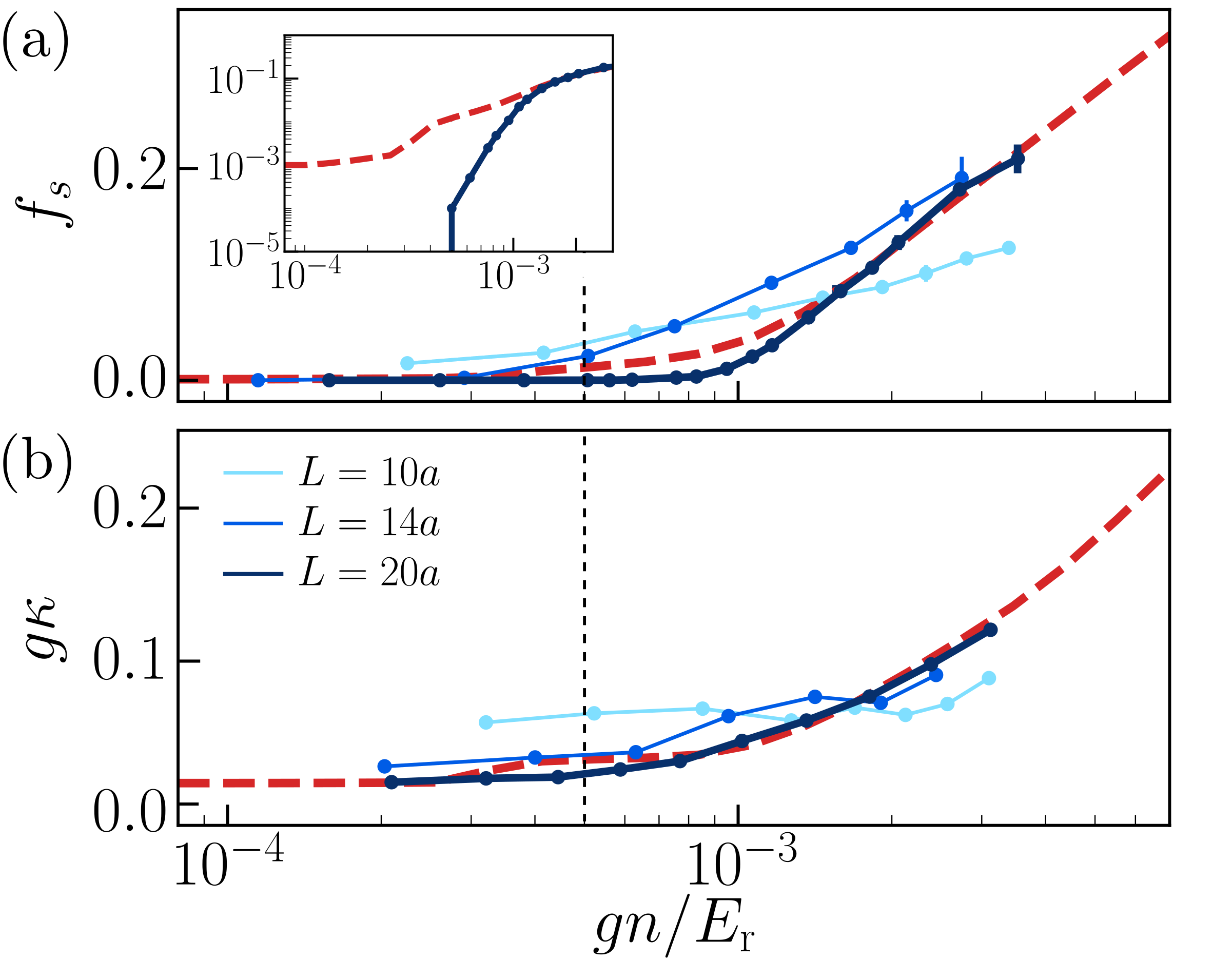}
\caption{\label{fig:fig2}
Interaction-driven BG-SF transition for $V_0=2.2\Er$.
(a)~SF fraction (semi log scale), with magnification of the critical region in inset (log-log scale), and
(b)~compressibility (semi log scale).
The dashed red lines show the mean field GPE results for a system size $L=20a$.
The blue points and solid lines show the QMC results for a fixed interaction parameter, $\tildegnot = 0.03$, and increasing system sizes $L/a = 10, 14, 20$ (from light to dark blue).
The dashed black line indicates the BG-SF critical point.
}
 \end{figure}

While these results are compatible with a SF-BG phase transition, the critical line is smoothed out by mean field effects, see Fig.~\ref{fig:fig2}(a) (dashed red line).
%
To locate the transition line accurately, we now turn to \textit{ab initio} quantum Monte Carlo (QMC) calculations. Our algorithm relies on the continuous space, path integral representation, simulating the exact Eq.~(\ref{eq:Hamiltonian}) Hamiltonian. The QMC configurations are efficiently sampled using the worm algorithm within the grand canonical ensemble~\cite{boninsegni2006,boninsegni2006b}, and we use a generalized interaction propagator applicable to both weak and strong interactions~\cite{note:SupplMat}.
We work at a vanishingly small temperature, $T=0.0025\Er/\kB$~\cite{note:temp}, and a fixed value of the interaction parameter, $\tildegnot=0.03$.
For such a small value, the $\mu$-dependent term in Eq.~(\ref{eq:coupling}) is negligible in our calculations and $\tildeg \simeq \tildegnot \ll 1$, corresponding to the weakly interacting regime.
For each value of the potential amplitude $V_0$, we then scan the chemical potential $\mu$ and determine the density $n$ from the statistics of the QMC world lines. The SF fraction is determined as $\fs=\Upsilon/n$, where the SF stiffness $\Upsilon$ is computed using the winding number estimator with periodic boundary conditions~\cite{ceperley1995}.

The QMC results for the SF fraction are plotted versus the mean field interaction parameter $gn$ on Fig.~\ref{fig:fig2}(a) for increasing system sizes (corresponding to darker blue lines).
For a large enough system, the QMC results fit well the mean field GPE prediction except in the critical region. While the GPE result is smooth, the QMC results show a sharp transition from the BG phase ($\fs=0$) to the SF phase ($\fs>0$), see magnification in log-log scale in the inset of Fig.~\ref{fig:fig2}(a).
Proceeding similarly for various amplitudes $V_0$ of the quasicrystal potential
and still $\tildegnot=0.03$,
we determine the exact SF-BG critical line shown on Fig.~\ref{fig:fig1} (see pink points; the dotted line is a guide to the eye). The QMC critical line fits well within the mean field crossover region.
Hence, although criticality requires truly many-body effects,
mean field calculations yield a reasonable estimate of the SF-BG transition.

We have also computed the compressibility $\kappa = \partial n / \partial \mu$ across the transition. Mean field and QMC results are plotted on Fig.~\ref{fig:fig2}(b),
showing an excellent agreement for sufficiently large systems.
In contrast to the SF fraction, the compressibility shows a smooth crossover from the BG phase to the SF phase which is consistent with the expectation that both are compressible phases.
More precisely, the compressibility is progressively suppressed by localization all the way from the SF limit to the BG limit, with no signature of the critical point.

\lettersection{Strong interactions}
%
 \begin{figure*}[t!]
         \centering
         \includegraphics[width = 2.0\columnwidth]{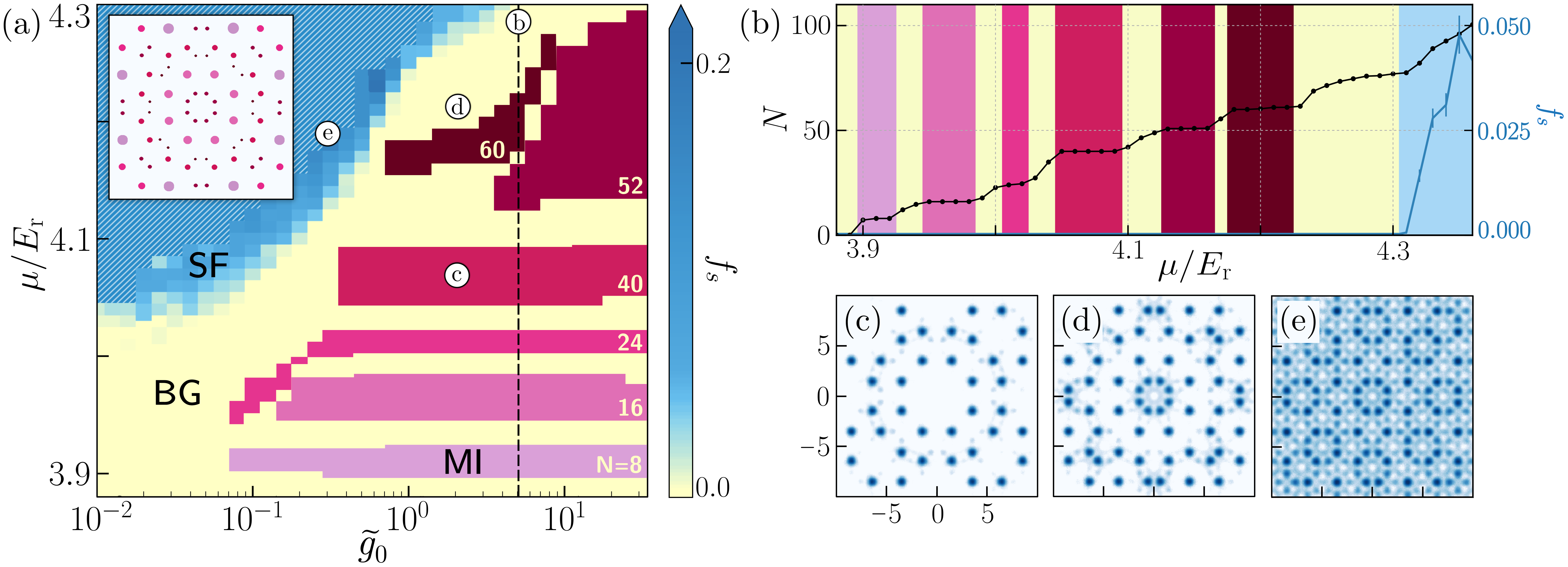}
\caption{\label{fig:fig3}
(a)~Phase diagram versus the interaction parameter $\tildegnot$ and the chemical potential $\mu$ for $V_0=2.5 \Er$ and a size $L=20a$ as found using QMC calculations. The SF fraction $\fs$ is shown in yellow-blue color scale. The purple-red regions indicate MI lobes with fillings $N=8, 16, 24, 40, 52, 72$ (from lighter to darker), respectively.
The Inset is a sketch of the potential wells, each colored according to the color of the MI lobe that fills it first.
(b)~Vertical cut in the phase diagram at $\tildegnot=5$ (dashed black line), showing the number of particles $N$ (black joint dots) and the superfluid fraction $\fs$ (blue joint dots) against the chemical potential $\mu$.
(c)-(e)~Density profiles in logarithmic color scale at the three points indicated on panel~(a), in, respectively, the $N=40$ MI, the BG, and the SF phases.
}
 \end{figure*}
%
We now turn to strongly interacting bosons.
In this case, the mean field GPE approach is no longer valid and we only rely on QMC calculations. We compute the superfluid fraction $\fs$ and the compressibility $\kappa$ as above, using the generalized interaction propagator~\cite{note:SupplMat}.
Figure~\ref{fig:fig3}(a) shows the phase diagram against the interaction parameter $\tildegnot$ and the chemical potential $\mu$ for the potential amplitude $V_0=2.5 \Er>\Vc$ and a square system of linear size $L=20a$.
For $\tildegnot \ll 1$, we recover the SF-BG phase transition discussed above. The critical point for $\tildegnot=0.01$ is found at $\muc \simeq 4.03 \Er$ and $g\nc \simeq 1.9 \times 10^{-3}\Er$, see Fig.~\ref{fig:fig3}(a) and the hollow pink points on Fig.~\ref{fig:fig1}.
This behavior persists up to $\tildegnot \sim 0.06$.
For stronger interactions, beyond mean field effects shift the critical point toward higher chemical potentials and, correspondingly, higher densities.
Moreover, Mott lobes open (MI, purple-red regions).
Figure~\ref{fig:fig3}(b) is a cut of the phase diagram at $\tildegnot=5$, see black dashed line on Fig.~\ref{fig:fig3}(a).
It shows a clear SF transition from $\fs=0$ to $\fs>0$ at $\muc \simeq 4.31\Er$.
In the insulating phase ($\fs=0$), we find a series of Mott plateaus characterized by a vanishing compressibility,
$\kappa = L^{-2} \partial N / \partial \mu = 0$.
We have checked that, consistently,
the fluctuations of the total number of particles $N$ exactly vanishes.
These Mott lobes emerge due to large repulsive interactions, which localize the particles in the deepest wells, thus forming an incompressible insulator with integer fillings. Because of the eightfold rotational symmetry of the quasicrystal potential, the total number of particles is usually a multiple of $8$,
namely $N=8$, $16$, $24$, $40$, see numbers on Fig.~\ref{fig:fig3}(a).
They progressively fill the lowest-energy potential wells, sketched with corresponding colors and decreasing sizes in the inset of Fig.~\ref{fig:fig3}(a).

Two exceptions are the $N=52$ and $N=60$ lobes. They feature $40$ particles localized in the $40$ deepest wells similar to the fourth lobe, plus additional particles in the next $24$ wells. Among these wells, $16$ come in $8$ pairs of close-by wells, corresponding to the small dots on the very top center of the inset of Fig.~\ref{fig:fig3}(a) and those obtained by successive rotations of angle $\pi/4$.
For the $N=60$ lobe, all these sites are populated by a particle while for the $N=52$ lobe, only one of the partners of each pair is populated while the other one is empty.
The remaining eight wells form the shortest eightfold ring in the center and one of every two wells contains a particle for both the $N=52$ and $N=60$ lobes.
For each lobe, there are several configurations corresponding to the choice of the filled and empty sites, and we find that the QMC density profile randomly blinks between them all, hence spontaneously breaking the eightfold rotation symmetry.
Similar symmetry breaking was previously found for long-range interactions~\cite{johnstone2019}. Here, we quite unexpectedly find it for short-range interactions.
We attribute this behavior to significant repulsion from the tails of the states bound in the nearest wells of the quasicrystal potential. Although their overlap is weak, strong-enough interactions prevent two particles from sitting even in different wells.

Figure~\ref{fig:fig3}(c)-(e) show the density profile for three selected points in the phase diagram, see Fig.~\ref{fig:fig3}(a). The first one corresponds to the $N=40$ MI with exactly one particle in the $40$ deepest wells.
The second one corresponds to a BG. It features $N = 56$ particles in $56$ distinct wells plus an incommensurate number of particles in the central eight wells. The latter form a small superfluid ring, which contributes a small finite compressibility but not global superfluidity. When the chemical potential increases or the interactions decrease, such local superfluids proliferate and finally merge. The third density plot corresponds to a global SF with delocalized particles.

\lettersection{Conclusions}
We have determined the quantum phase diagram of weakly and strongly interacting 2D bosons in a shallow quasicrystal potential.
The SF, BG, and MI quantum phases can be observed in current generation experiments with ultracold atoms using a combination of interference, spectroscopy, and transport measurements~\cite{derrico2014,gori2016,sbroscia2020}.
We have considered the same eightfold potential as in Refs.~\cite{viebahn2019,sbroscia2020} but we expect our results to qualitatively hold also for other quasicrystalline potentials~\cite{guidoni1997,lsp2005} as well as other configurations designed for photonic systems in the nonlinear regime~\cite{valy2013}.
Control of the 2D interaction in the range $0.05 \lesssim \tildeg \lesssim 3$ has been demonstrated in Ref.~\cite{ha2013}. It is sufficient to observe the phase diagram of Fig.~\ref{fig:fig3}(a) and in particular the still elusive BG phase in 2D.
Using sufficiently large samples in box-shaped traps could, for instance, yield further insight to the transitions discussed in this work, including critical exponents.

The use of shallow quasiperiodic potentials as considered in this Letter is promising for overcoming stringent temperature effects in the observation of the BG phase, as recently shown in 1D~\cite{yao2019}.
Yet,
finite temperatures show richer behavior in 2D compared to 1D, in particular topological phase transitions.
For instance, in a uniform 2D Bose gas, the SF-to-normal fluid transition is of the Berezinskii-Kosterlitz-Thouless type, at some critical temperature~\cite{berezinskii1972,kosterlitz1973,hadzibabic2006,desbuquois2012,plisson2011}. 
While disorder with short-range correlations does not affect this transition~\cite{harris1974,carleo2013}, to our knowledge, the effect of long-range quasiperiodic order remains an open question.

\begin{acknowledgments}
We thank Markus Holzmann and H\'el\`ene Perrin for discussions about 2D scattering theory, and the CPHT computer team for valuable support.
This research was supported by the Paris region DIM-SIRTEQ.
The numerical calculations were performed using HPC resources from GENCI-CINES (Grant 2019-A0070510300) and make use of the ALPS scheduler library and statistical analysis tools~\cite{troyer1998,ALPS2007,ALPS2011}.
\end{acknowledgments}


 \renewcommand{\theequation}{S\arabic{equation}}
 \setcounter{equation}{0}
 \renewcommand{\thefigure}{S\arabic{figure}}
 \setcounter{figure}{0}
 \renewcommand{\thesection}{S\arabic{section}}
 \setcounter{section}{0}
 \onecolumngrid  
     
 
 \newpage
 
{\center \bf \large Supplemental Material for \\}
 {\center \bf \large Strongly-Interacting Bosons in a Two-Dimensional Quasicrystal Lattice\\ \vspace*{1.cm}
 }

In this supplemental material, we provide details about the single-particle localization properties of the 2D quasicrystal potential (Sec. S1), 2D scattering theory (Sec. S2), and quantum Monte Carlo (QMC) calculations, including the derivation of the complete interaction propagator (Sec. S3).

\begin{figure}[b!]
  \centering
  \includegraphics[width = 1\columnwidth]{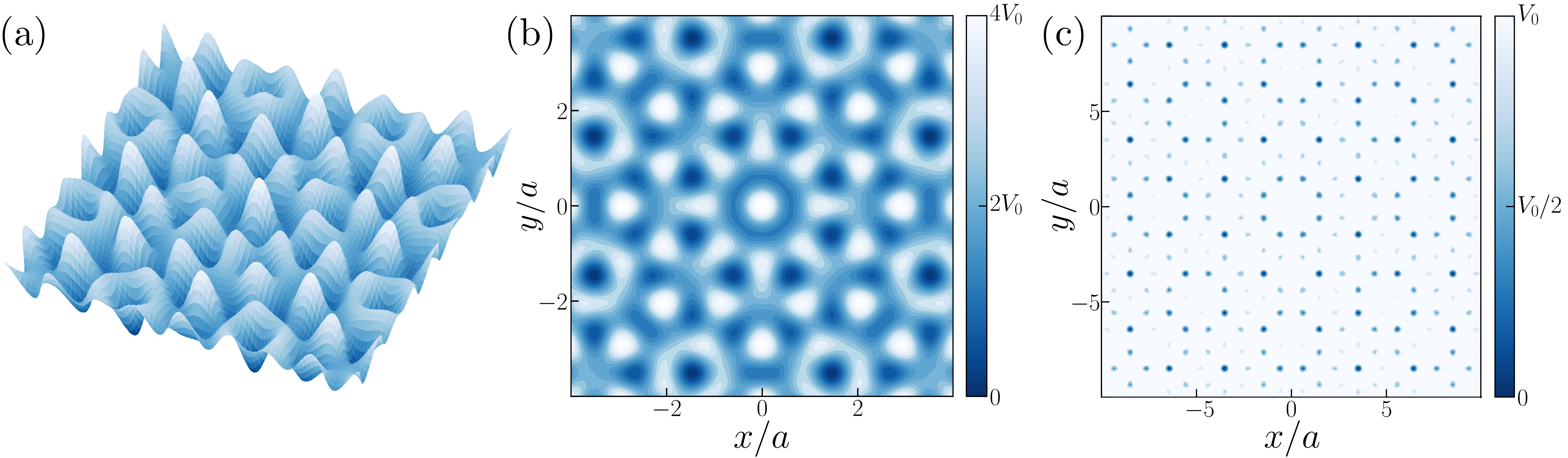}
\caption{\label{fig:figs1}
Eightfold quasicrystal potential of Eq.~(2) in the main paper.
(a)~3D plot for the size $L = 8a$.
(b)~Same as~(a) shown from the top.
(c)~2D plot for size $L=20a$ with a truncation of the colorscale at $V=V_0$.
The values of the potential between $V_0$ and $4V_0$ thus appear in white, hence enlightening the potential wells (dark blue).
}
\end{figure}

\section{Single-particle problem}

In this section, we discuss the single-particle localization transition and the critical exponent for the two-dimensional (2D) quasicrystal potential of Eq.~(2) in the main paper, using the same approach as in Refs.~\cite{yao2019,yao2020}. The quasicrystal potential is shown in Fig.~\ref{fig:figs1}. The two left panels show the potential $V(\rr)$, with $\rr$ the 2D vector position, for a square lattice of size $L \times L$ with $L = 8a$,
as a 3D colorplot in~(a) and as a 2D plot in~(b).
Panel~(c) shows $V(\rr)$ truncated at $V = V_0$ for $L=20a$, hence highlighting the potential wells.

To study the single-particle localization properties of this quasicrystal potential, we solve the Hamiltonian
\begin{equation}\label{eq:Hamiltonian0}
  \hat{h} = -\frac{\hbar^2}{2m} \nabla^2 + V(\hat{\rr}),
\end{equation}
that is the Hamiltonian~(1) of the main paper for a single particle, using exact diagonalization. We compute the inverse participation ratio ($\IPR$) of the $n$-th eigenstate $\psi_n$,
\begin{equation}
  \IPR_n = \frac{ \int \dd \rr |\psi_n(\rr)|^4}{\left( \int \dd \rr |\psi_n(\rr)|^2 \right)^2}.
\end{equation}
In 2D, it scales as $\IPR_n \sim 1/L^2$ for an extended state and as $\IPR_n \sim 1$ for a localized state. 

\begin{figure}[t!]
  \centering
  \includegraphics[width = 1\columnwidth]{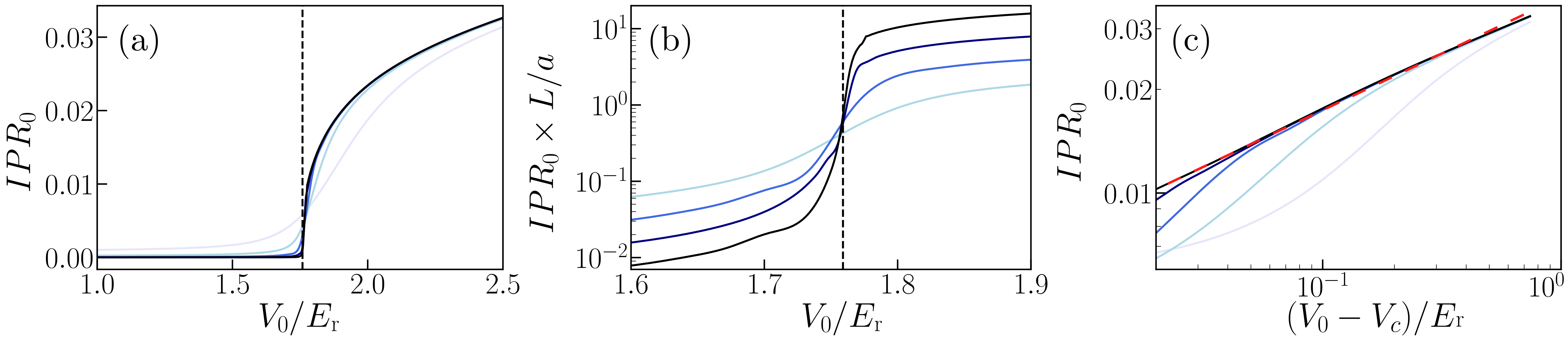}
\caption{\label{fig:figs2}
Critical potential and critical exponent for the single-particle problem in the quasicrystal potential of Eq.~(2) in the main paper.
(a)~IPR of the ground state, $\IPR_0$, versus the quasicrystal amplitude $V_0/\Er$. Darker lines correspond to increasing system sizes:
$L/a = 16$ (light grey), $32$ (light blue), $64$ (blue), $128$ (dark blue), and $256$ (black).
(b)~Rescaled IPR of the ground state, $\IPR_0 \times L/a$, using the same data as in panel~(a).
(c)~$\IPR_0$, in log-log scale, for a potential amplitude above the critical value $\Vc \simeq 1.76(1)$. The dashed red line is a fit to the black line, corresponding to the largest system. It yields the critical exponent $\nu \simeq 0.33(1)$.
}
\end{figure}

Figure \ref{fig:figs2}(a) shows a finite-size scaling of the IPR of the  ground state, $\IPR_0$, against the potential amplitude $V_0/\Er$ for increasingly large systems, from $L=16a$ (light grey) to $L=256a$ (black). Increasing the potential amplitude $V_0$, a sharp transition is found from the extended phase with vanishingly small $\IPR$ to the localized phase with finite $\IPR$.
To locate the critical point accurately, we rescale the $\IPR$ by the size of the system $L$. It yields a new quantity, $\IPR \times L/a$, that scales as $1/L$ for an extended phase and $L$ for a localized phase. Figure~\ref{fig:figs2}(b) shows this rescaled $\IPR$ in log scale in a range of potential amplitudes nearby to the critical potential. In this figure, all the lines corresponding to increasing system sizes cross at a nearly single point thus indicating the critical point. The critical potential is calculated as such and yields $\Vc/\Er \sim 1.76(1)$ in good agreement with the result of Ref.~\cite{szabo2020} found using another approach (ground-state curvature). 

We have also studied the critical exponent $\nu$ of the potential, defined as $\IPR_0 \sim (V_0 - V_c)^\nu$. In Figure~\ref{fig:figs2}(c), we plot the $\IPR_0$ against the potential amplitude $(V_0 - V_c)/\Er$ in log-log scale, for the same system sizes as in the panels~(a) and (b). For a large-enough system, we find a straight line in log-log scale, confirming the scaling $\IPR \propto (V_0 - V_c)^\nu$. Note that all the finite-size simulations converge towards the same straight line for large enough potential amplitudes. The red dashed line shows a linear fit of the largest size (black line). It yields the critical exponent of $\nu = 0.33(1)$. Remarkably, this critical exponent is similar to that found in Ref~\cite{yao2019} for 1D quasiperiodic potentials. 

\section{Scattering theory in quasi-2D Bose gases}
In Refs.~\cite{petrov2000a,petrov2001}, Petrov~\etal\ have solved the scattering problem of two bosons interacting via a 3D short-range interaction potential in a quasi-2D geometry. The interaction strength in free 3D space is characterized by the scattering length $\asc$.
The bosons are confined by a harmonic trap in the direction orthognonal to the 2D plane of interest, with the angular frequency $\omega_\perp$.
Assuming $\hbar\omega_\perp \gg \kB T, \mu$, the dynamics is frozen to zero point transverse oscillations and the scattering amplitude in this channel takes the form of its purely-2D counterpart, with the 2D scattering length~\cite{pricoupenko2007}
\begin{equation}\label{eq:a2d-a3dSM}
\atwod \simeq 2.092\aho \mathrm{exp}\left( -\sqrt{\frac{\pi}{2}} \frac{\aho}{\asc} \right),
\end{equation}
with  $\aho = \sqrt{\hbar/m\omega_\perp}$ the width of the transverse harmonic oscillator, \ie\ Eq.~(3) of the main paper.
Moreover, they find the mean field coupling constant
\begin{equation}\label{eq:coupling-petrov}
\tildeg \simeq \frac{2\sqrt{2\pi}}{\aho/\asc+{1/\sqrt{2\pi} \ln (1/\pi q^2 \aho^2)}},
\end{equation}
with $q=\sqrt{2m|\mu|/\hbar^2}$ the quasi-momentum.

\bigskip
\textit{Elimination of the transverse degrees of freedom.~---~}
In our mean field and quantum Monte Carlo calculations, the transverse direction is integrated out and we work in 2D. It is thus convenient to eliminate the transverse degrees of freedom from Eq.~(\ref{eq:coupling-petrov}). Using Eq.~(\ref{eq:a2d-a3dSM}), we substitute the first term in the denominator by $\aho/\asc=\sqrt{2/\pi}\ln\left(2.092\aho/\atwod\right)$. Then, writing $q$ as a function of $\mu$ and using logarithmic identities, we find
\begin{equation}\label{eq:couplingSM}
\tildeg \simeq \frac{4\pi}{2\ln(a/\atwod) + \ln\left(\Lambda \Er/\mu\right)},
\end{equation}
with $\Lambda \simeq 2.092^2/\pi^3 \simeq 0.141$
and $\Er = \pi^2\hbar^2/2ma^2$ the recoil energy.
Note that $a$ may be seen as an arbitrary length and one can eliminate it from Eq.~(\ref{eq:couplingSM}) using logarithmic identities.
Here, we take $a$ to be the lattice spacing of the quasicrystal potential.
Equation~(\ref{eq:couplingSM}) is the same Eq.~(4) of the main paper with the interaction parameter
\begin{equation}\label{eq:coupling0SM}
\tildegnot = \frac{2\pi}{\ln(a/\atwod)}
\end{equation}
used in our work.
It is a convenient form in our work for two reasons.
Firstly, all the transverse degrees of freedom have been eliminated, in particular the transverse length $l_\perp$.
Secondly, we separated out the contributions of the parameter characterizing the interaction strength, \ie\ $\atwod$ or equivalently $\tildegnot$, and from the chemical potential $\mu$.

\bigskip
\textit{Quasi-2D versus purely-2D regimes.~---~}
We now discuss the relevant scattering regimes in ultracold-atom experiments.
Here we rely on Eqs.~(\ref{eq:a2d-a3dSM}) and (\ref{eq:coupling-petrov}) rather than Eqs.~(\ref{eq:couplingSM}) and (\ref{eq:coupling0SM}).

\bigskip
For intermediate confinement strengths such that
\begin{equation}\label{eq:quasi2D}
\asc \ll \aho
\qquad \textrm{and} \qquad
\mu \ll \hbar\omega_\perp,
\end{equation}
the dynamics is frozen to zero point oscillations but the scattering conserves its 3D character~\cite{petrov2000a,petrov2001}. Except in extreme cases, the logarithmic term in the denominator of Eq.~(\ref{eq:coupling-petrov}) is negligible and one finds
\begin{equation}\label{eq:quasi2d-g}
\tildeg=\frac{2\sqrt{2\pi}\athreed}{\aho}
=\frac{2\pi}{\ln \left( \frac{2.092\aho}{\atwod} \right)},
\end{equation}
where we have used Eq.~(\ref{eq:a2d-a3dSM}).
This regime, known as the \textit{quasi-2D regime} is the most usual one in ultracold-atom experiments, see for instance Refs.~\cite{hadzibabic2006,zhang2012,de2016,sbroscia2020}.
Since $\asc \ll \aho$, the quasi-2D Bose gas is always in the weakly interacting regime, $\tildeg \ll 1$.
Note that the quasi-2D regime is also characterized by $\atwod \ll \aho \ll \lambda_{T}, \xi$. Owing to the logarithm, however, it requires that $\atwod$ is significantly much smaller than $l_\perp$.

\bigskip
For a stronger confinement,
\begin{equation}\label{eq:towardspurely2D}
\aho \lesssim \asc
\qquad \textrm{and} \qquad
\mu \ll \hbar\omega_\perp,
\end{equation}
the logarithmic term in the denominator of Eq.~(\ref{eq:coupling-petrov}) may dominate and
one finds
\begin{equation}\label{eq:towardspurely2Dg}
\tildeg \simeq \frac{4\pi}{\ln\left(\frac{\hbar^2}{2\pi m \mu \aho^2}\right)}.
\end{equation}
It is then possible to reach the strongly-interacting regime with $\tildeg \sim 1$ for moderately small chemical potentials. For instance, values up to $\tildeg \simeq 3$ have been reported in Ref.~\cite{ha2013}.

\bigskip
For very strong confinement strengths, 
\begin{equation}\label{eq:purely2D}
\aho \ll \asc
\qquad \textrm{and} \qquad
\mu \ll \hbar\omega_\perp,
\end{equation}
the 2D scattering length saturates,
\begin{equation}\label{eq:a2d-a3dSMpurely}
\atwod \simeq 2.092\aho,
\end{equation}
see Eq.~(\ref{eq:a2d-a3dSM}).
This regime is known as the \textit{purely-2D regime}.
In the ultra-dilute limit, $n\atwod^2 \ll 1$, we may use the equation of state of the purely 2D Bose gas~\cite{schick1971},
\begin{equation}\label{eq:EOS-2d}
\mu= \frac{4\pi \hbar^2 n / m}{\mathrm{ln} (1/n\aTwoD^2)}.
\end{equation}
One then finds
$$
\ln\left(\frac{\hbar^2}{2\pi m \mu \aho^2}\right) \simeq \ln\left(1/n\atwod^2\right) +  \ln \ln\left(1/n\atwod^2\right) + \textrm{cst}.
$$
Inserting this expression into Eq.~(\ref{eq:towardspurely2Dg}), one finds
\begin{equation}\label{eq:purely2d-g}
\tildeg \simeq  \frac{4\pi}{\ln (1/n\atwod^2)}
\end{equation}
up to logarithmic accuracy. It is consistent with Eq.~(\ref{eq:EOS-2d}) and $\mu \simeq \tildeg \hbar^2 n/m$.

\bigskip
Note that here we have defined the terms quasi-2D and purely-2D according to the respective values of the 3D scattering length $\asc$ and the oscillation length $\aho$, similarly as in Refs.~\cite{pricoupenko2007,bloch2008,mora2003,hadzibabic2011,carleo2013,ha2013,de2016} for instance. 
Other authors define it by comparing the temperature $\kB T$ with the  energy scale $\hbar\omega_\perp$, see for instance Refs.~\cite{petrov2000a,petrov2000b,petrov2001}.

\section{Quantum Monte Carlo calculations}\label{sec:propagator}
In this section, we outline the path integral Monte Carlo (PIMC) approach used in the main paper. In particular, we show how to derive a complete 2D interaction propagator valid in any regime of interactions.

\subsection{Generalities}
We consider a system of $N$ identical bosons
governed by the Hamiltonian $\hat{H}$.
In the canonical ensemble at temperature $T$, the expectation value of an observable $\hat{A}$ reads as
\begin{equation}\label{eq:avA}
	\langle A \rangle = \frac{\Tr \left( \e^{-\beta \hat{H}} \hat{A} \right)}{\Tr \left( \e^{-\beta \hat{H}}\right)},
\end{equation}
where $\beta = 1/\kB T$ and $\Tr(\hat{X}) = 1/N! \sum_{\sigma \in \mathfrak{S}_N} \int \dd \vect{R} \bra{\sigma \cdot \vect{R}} \hat{X} \ket{\vect{R}}$. In this expression of the trace, the particle positions are contained in a collective variable, $\vect{R} = (\vect{r}_1, \vect{r}_2, \ldots \vect{r}_N)$, and $\sigma \cdot \vect{R}= (\vect{r}_{\sigma(1)}, \vect{r}_{\sigma(2)}, \ldots \vect{r}_{\sigma(N)})$, where $\sigma$ spans the $N!$ permutations of indices.
To derive an expression suitable for numerical estimation, the exponential terms in Eq.~\eqref{eq:avA} are split into $J$ factors of imaginary time $\epsilon = \beta/J$ and completeness relations are inserted at every time slice, such that the numerator of Eq.~\eqref{eq:avA} becomes
\begin{equation}\label{eq:TrebHA}
	\Tr \left( \e^{-\beta \hat{H}} \hat{A} \right) = \frac{1}{N!} \sum_{\sigma \in \mathfrak{S}_N} \int \dd \vect{R}_0 \dd \vect{R}_1 \ldots \dd \vect{R}_{J-1} \bra{\sigma \cdot \vect{R}_0} \e^{-\epsilon \hat{H}} \ket{\vect{R}_{J-1}} \ldots \bra{\vect{R}_1} \e^{-\epsilon \hat{H}} \hat{A}\ket{\vect{R}_0}.
\end{equation}
The denominator of Eq.~\eqref{eq:avA} is the same as Eq.~(\ref{eq:TrebHA}) with $\hat{A}=1$.
A \textit{configuration} $\mathcal{C}$ is the set of positions of the particles at every time slice $ \left( \vect{R}_0, \ldots,\vect{R}_{J-1}, \vect{R}_J = \sigma \cdot \vect{R}_0 \right)$ where $\vect{R}_J$ is a permutation of the initial positions due to the trace. The integrals and sums in Eq.~\eqref{eq:TrebHA} are thus replaced by an integration over all the configurations $\mathcal{C}$. The \textit{path-integral estimator} for the observable $\hat{A}$ is defined as $\mathcal{A}(\mathcal{C}) \equiv \bra{\vect{R}_1} \e^{-\epsilon \hat{H}} \hat{A} \ket{\vect{R}_0} / \bra{\vect{R}_1} \e^{-\epsilon \hat{H}} \ket{\vect{R}_0}$. In practice, using the invariance under cyclic permutation of the trace in Eq.~(\ref{eq:TrebHA}), one can make the operator $\hat{A}$ appear between any pair of consecutive positions $\vect{R}_j$ and $\vect{R}_{j+1}$. Therefore, to take full advantage of the information contained in a given configuration $\mathcal{C}$, the path-integral estimator is numerically calculated as the average over all pairs of consecutive positions. With these notations, the expectation value of $\hat{A}$ can be rewritten as
\begin{equation}\label{eq:avAbis}
  \langle A \rangle = \int \dd \mathcal{C}\, \pi(\mathcal{C}) \mathcal{A}(\mathcal{C})
\end{equation}
where
\begin{equation}\label{eq:probapi}
\pi(\mathcal{C}) = \frac{\bra{\sigma \cdot \vect{R}_0} \e^{-\epsilon \hat{H}} \ket{\vect{R}_{J-1}} \ldots \bra{\vect{R}_1} \e^{-\epsilon \hat{H}} \ket{\vect{R}_0}}{\int \dd \mathcal{C} \bra{\sigma \cdot \vect{R}_0} \e^{-\epsilon \hat{H}} \ket{\vect{R}_{J-1}} \ldots \bra{\vect{R}_1} \e^{-\epsilon \hat{H}} \ket{\vect{R}_0}}
\end{equation}
is the statistical weight of configuration $\mathcal{C}$. Note that the normalization condition is $\int \dd \mathcal{C}\, \pi(\mathcal{C}) = 1$.
As explicitely shown by Eqs.~\eqref{eq:avAbis} and \eqref{eq:probapi}, the many-body propagator $\rho(\vect{R}, \vect{R}', \epsilon) = \bra{\vect{R}'} \e^{-\epsilon \hat{H}} \ket{\vect{R}}$ plays a central role in the evaluation of $\langle A \rangle$.
Its derivation are discussed in the next sections.

\subsection{Pair-product approximation}
We now consider a system governed by a Hamiltonian with two-body interactions, given by
\begin{equation}\label{eq:manybodyhamil}
  \hat{H} = \sum_j h(\hat{\rr}_j) + \sum_{j < k} U(\hat{\rr}_j - \hat{\rr}_k).
\end{equation}
where $h(\hat{\rr}_j)$ is the single-particle Hamiltonian of particle $j$ (including the kinetic and potential energies), and $U(\hat{\rr}_j - \hat{\rr}_k)$ is the interaction Hamiltonian between particles $j$ and $k$.
Due to translation invariance, the interaction term only depends on the relative position of the two particles $j$ and $k$, \ie\ $\vect{r}_{jk} = \vect{r}_j - \vect{r}_k$,
and it is convenient to use the interacting Hamiltonian of the reduced particle~\cite{yangqian15}
\begin{equation}\label{eq:twobodyinteraction}
  H^\text{rel}(\hat{\rr}_{jk}) = H^\text{rel}_0(\hat{\rr}_{jk}) + U(\hat{\rr}_{jk}),
\end{equation}
where $H^\text{rel}_0(\hat{\rr}_{jk}) = \hbar^2 \hat{\mathbb{\nabla}}_{\rr_{jk}}^2/m$ is the noninteracting Hamiltonian (kinetic term) of the reduced particle, the mass of which is $m/2$.
Inserting Eq.~(\ref{eq:twobodyinteraction}) into Eq.~(\ref{eq:manybodyhamil}) and using the
pair-product approximation, the many-body propagator
then reads as
\begin{equation}
  \rho(\mathbf{R}, \mathbf{R'}, \epsilon) \approx \prod_{j=1}^N \rho_1(\mathbf{r}_j, \mathbf{r}'_j, \epsilon) \prod_{j<k}^N \frac{\rho^{\text{rel}}(\mathbf{r}_{jk}, \mathbf{r}'_{jk}, \epsilon)}{\rho^{\text{rel}}_0(\mathbf{r}_{jk}, \mathbf{r}'_{jk}, \epsilon)},
\end{equation}
where $\rho_1$ is the one-body density matrix of the noninteracting problem,
and
\begin{equation}
\rho^{\text{rel}} (\mathbf{r}_{jk}, \mathbf{r'}_{jk}, \epsilon)
= \bra{\mathbf{r'}_{jk}} \e^{-\epsilon {\hat{H}^\text{rel}}} \ket{\mathbf{r}_{jk}}
\end{equation}
is that of the relative particle, $\rho^{\text{rel}}_0 (\mathbf{r}_{jk}, \mathbf{r'}_{jk}, \epsilon)$ being its counterpart in the absence of interactions.

\subsection{Two-dimensional interaction propagator}
To derive the relative-particle propagator $\rho^{\text{rel}}(\mathbf{r}, \mathbf{r}', \epsilon)$, we consider two particles of mass $m$ and relative position $\mathbf{r} = \mathbf{r}_1 - \mathbf{r}_2$.
Due to global rotational invariance, the eigenstates of energy $E_k = \hbar^2 k^2 /m$
may be written in the form $\psi_{kl}(r,\theta) = R_{kl}(r)\Theta_l(\theta)$,
where the quantities $\Theta_l(\theta) = \e^{il\theta}/\sqrt{2\pi}$ are the eigenstates of the in-plane angular momentum operator, with $l \in \mathbb{Z}$ the corresponding quantum number.
The propagator then reads as
\begin{equation}\label{eq:propag}
    \rho^{\text{rel}}(\mathbf{r},\mathbf{r'}, \epsilon) = \sum_l \int  \dd k\ \psi_{kl}^*(\mathbf{r'}) \e^{-\epsilon k^2} \psi_{kl}(\mathbf{r}).
\end{equation}
Plugging the expression $\psi_{kl}(r,\theta) = R_{kl}(r)\Theta_l(\theta)$ into the 2D Schr\"odinger equation yields
\begin{equation}\label{eq:BesselWithU}
  \frac{\dd^2 R_{kl}}{\dd r^2} + \frac{1}{r} \frac{\dd R_{kl}}{\dd r} + \left(k^2 - \frac{l^2}{r^2}\right) R_{kl}  = \frac{mU(r)}{\hbar^2} R_{kl}.
\end{equation}

For noninteracting particles, $U(r)=0$, Eq.~(\ref{eq:BesselWithU}) is known as the Bessel equation, the general solution of which reads as
\begin{equation}\label{eq:radialbessel}
  R_{kl}(r) = \mathcal{C}_{kl} \left[\cos(\delta_{kl}) J_l(kr) - \sin(\delta_{kl}) Y_l(kr) \right]
\end{equation}
where $J_l$ and $Y_l$ are Bessel functions of the first and second kind respectively, $\mathcal{C}_{kl}$ is a normalization constant and $\delta_{kl}$ is a phase shift. In the noninteracting case, the set of solutions with a nonzero contribution of $Y_l$ are expected to vanish due to their divergence at $r=0$, thus yielding $\delta_{kl} = 0$.
In addition, the normalization and completeness relations $\int \dd \rr \int \dd k\ \psi_{kl}^*(\mathbf{r}) \psi_{kl}(\mathbf{r}) = 1$ and $\int \dd k\ \psi_{kl}^*(\mathbf{r}') \psi_{kl}(\mathbf{r}) = \delta(\rr - \rr')$ yield $\mathcal{C}_{kl} = \sqrt{k}$.
The noninteracting wavefunction is thus given by $\psi_{kl}(r, \theta) = \sqrt{k/2\pi} J_l(kr) \e^{il\theta}$. Inserting this formula into the Eq.~(\ref{eq:propag}) and using
the following identities on Bessel functions,
\begin{equation}
  \int k \e^{-\epsilon k^2} J_l(kr) J_l(kr') \mathrm{d}k = \frac{1}{2\epsilon} \exp \left(-\frac{r^2+r'^2}{4\epsilon}\right) I_l \left( \frac{rr'}{2\epsilon} \right)
\qquad \textrm{and} \qquad
 \sum_l I_l(x) t^l = \exp \left(\frac{1}{2}x \left(t + \frac{1}{t}\right) \right),
\end{equation}
we then recover the well-known free-particle propagator,
\begin{equation}\label{eq:noninteractingpropag}
    \rho^{\text{rel},0}(\mathbf{r},\mathbf{r'}, \epsilon)
= \frac{1}{4\pi\epsilon} \e^{-\frac{(\mathbf{r}-\mathbf{r}')^2}{4\epsilon}}.
\end{equation}

The full 2D interacting problem is generally difficult to solve. For a short-range interaction potential in the $s$-wave scattering limit, with 2D scattering length $\atwod$, however, 
the wavefunction is expected to keep its noninteracting form when the particles are sufficiently far apart.
Equation~(\ref{eq:radialbessel}) still holds but with a nonzero phase shift $\delta_{kl}$, which hence accounts for the short-range interactions.
In the $s$-wave scattering limit, the contributions with $l \neq 0$ are neglected and the phase shift is found from the boundary condition at $r=0^+$~\cite{khuri2009}.
It yields $t_k \equiv \tan \delta_{k0} = \pi / 2 \ln(\eta k \atwod) $, where $\eta = \e^\gamma/2 \approx 0.89054$ and $\gamma$ is Euler's constant~\cite{whitehead2016}.
Using the eigenbasis decomposition, Eq.~(\ref{eq:propag}),
the 2D interaction propagator reads
\begin{equation}\label{eq:propagator}
  \begin{split}
    \rho^{\text{rel}}(\vect{r},\vect{r'}, \epsilon)
    = \rho^{\text{rel},0}(\vect{r},\vect{r'}, \epsilon)
    & -\frac{1}{2\pi} \int \dd k\ k \e^{-\epsilon k^2} \frac{t_k^2}{1+t_k^2} J_0(kr)J_0(kr') \\
    & -\frac{1}{2\pi} \int \dd k\ k \e^{-\epsilon k^2} \frac{t_k}{1+t_k^2} \left[J_0(kr)Y_0(kr') + J_0(kr')Y_0(kr) \right] \\
    & + \frac{1}{2\pi} \int \dd k\ k \e^{-\epsilon k^2} \frac{t_k^2}{1+t_k^2} Y_0(kr)Y_0(kr').
  \end{split}
\end{equation}

In the numerical implementation of the worm-algorithm QMC, the pair-propagator $\rho^{\text{rel}}(\vect{r},\vect{r'}, \epsilon)$ in Eq.~(\ref{eq:propagator}) is preprocessed and tabulated for a grid-like set of position values.
The actual pair-propagator used in the simulations in then evaluated from the cubic interpolation of these tabulated values.

\begin{figure}[t]
  \centering
  \includegraphics[scale=1,width=0.75\textwidth]{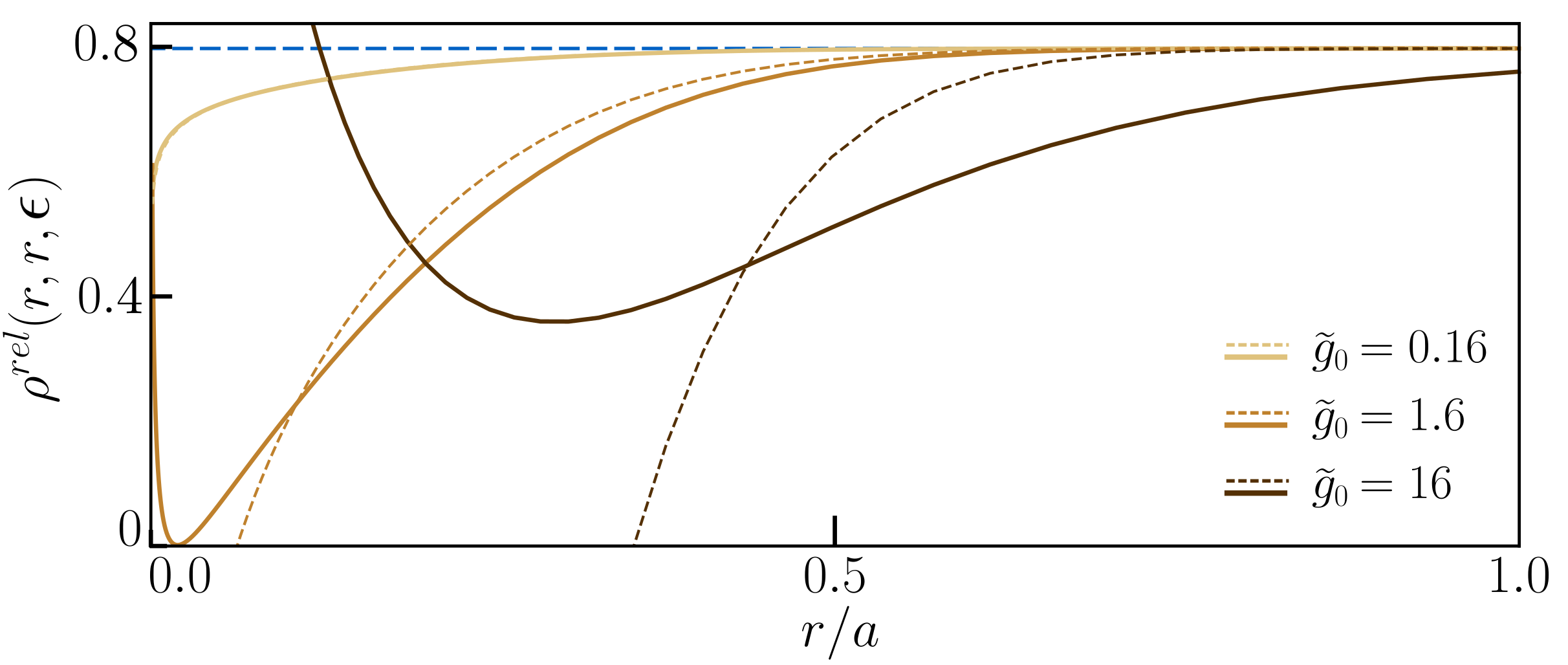}
  \caption{\label{fig:propagator}
  \small Diagonal term of the generalized interaction propagator (solid lines) and of the weak-interaction propagator as used in Ref.~\cite{carleo2013} (dashed lines), for three values of the interaction strength $\tildegnot$ and for $\epsilon = 0.1$. The lighter solid and dashed lines correspond to the weakly-interacting regime where both propagators match. The blue dashed line gives the noninteracting limit with $\rho^{\text{rel}, 0}(\vect{r},\vect{r}, \epsilon) = (4\pi\epsilon)^{-1}$. For each line, the corresponding values of 2D scattering length are
$\atwod / a \simeq 9 \times 10^{-18}$ (yellow, $\tildegnot=0.16$),
$\atwod / a \simeq 2 \times 10^{-2}$ (orange, $\tildegnot=1.6$),
and $\atwod / a \simeq 7 \times 10^{-1}$ (brown, $\tildegnot=16$).
  }
\end{figure}

Figure~\ref{fig:propagator} shows the diagonal term of the interaction propagator $\rho^{\text{rel}}(\vect{r},\vect{r}, \epsilon)$ for
various interaction strengths from weak to strong,
$\tildegnot = 0.16$ (yellow), for $\tildegnot = 1.6$ (orange) and for $\tildegnot = 16$ (brown).
We recall that the interaction parameter is
\begin{equation}\label{eq:coupling0SM}
\tildegnot = \frac{2\pi}{\ln(a/\atwod)},
\end{equation}
see Eq.~(5) of the main paper, which yields the relation
\begin{equation}\label{eq:tkVSg}
\frac{1}{t_k} + \frac{4}{\tildegnot} = \frac{2}{\pi}\ln(\eta ka).
\end{equation}
The diagonal terms of the propagator correspond to the probability of finding the two interacting particles at a distance $r$, while the nondiagonal terms give the probability for the two particles to jump from distance $r$ to $r'$. The diagonal terms thus give a rough idea of how terms with $r \approx r'$ behave. Figure~\ref{fig:propagator} shows that, as expected, the propagator increases with the distance between particles for $r>\atwod$ and that it converges to the noninteracting value (blue dashed line) for $r \gg \atwod$. However, the propagator diverges for $r \ll \atwod$ due to the nonexact treatment of the interaction term in the short-range region. To remedy this problem in numerical simulations, we chose to cut the propagator for radii below the scattering length,
\ie~$\rho^{\text{rel}}(\vect{r},\vect{r'}, \epsilon) = 0$ if $r < \atwod$ or $r' < \atwod$.
In practice, it removes the pairs of particles at a distance smaller than the 2D scattering length. It sets the validity condition of our simulations to $n\atwod^2 < 1$.
This condition is fulfilled for all our calculations.
Note also that our algorithm works in the grand-canonical ensemble. It allows for fluctuations in the total number of particles so that the averaged density is maintained in the simulations and does not drift to zero owing to particle removals.

\subsection{Limit of the propagator for weak interactions}
In the weakly-interacting regime, $\tildeg \simeq \tildegnot \ll 1$, Eq.~(\ref{eq:tkVSg}) can be simplified to $-t_k \approx \tildeg/4 \ll 1$ by neglecting the logarithmic correction.
In this regime, the interacting propagator, Eq.~(\ref{eq:propagator}) is dominated by the lowest (first)-order term in $t_k$ and we find
\begin{equation}
    \rho^{\text{rel}}(\vect{r},\vect{r'}, \epsilon) = \rho^{\text{rel},0}(\vect{r},\vect{r'}, \epsilon) + \frac{\tildeg}{8\pi} \int  \dd k\ k \e^{-\epsilon k^2} \left[J_0(kr)Y_0(kr') + J_0(kr')Y_0(kr) \right],
\end{equation}
which is the propagator as used in Ref.~\cite{carleo2013}.
The diagonal term of the weak-interaction propagator is shown on 
Fig.~\ref{fig:propagator} as dashed lines for the same values.
For the lowest value, $\tildegnot = 0.16$, the two propagators match perfectly (the dashed line, on top of the solid line, is not visble on the plot).
For higher values, $\tildegnot = 1.6$ and $\tildegnot = 16$, however, the two propagators exhibit significant differences, which impact the numerical results.
In particular, the full propagator has a richer behavior at large $r$.
While the weakly-interacting propagator quickly converges to the value of the non-interacting case at $1/4\pi\epsilon$, the full propagator shows deviation on a wider range.


\begin{thebibliography}{10}
\bibitem{note:temp}
This temperature is much smaller than any relevant temperature scale in the
  calculations. Moreover, we have checked that further decreasing the
  temperature does not significantly alter the results.
\end{thebibliography}
\end{document}